# Étude expérimentale et numérique des transferts de chaleur en convection naturelle le long de parois verticales épaisses rayonnantes fortement chauffées

**Thierry DE LAROCHELAMBERT**

Laboratoire Gestion des Risques et Environnement, Université de Haute-Alsace
25, rue de Chemnitz - 68200 MULHOUSE

**Résumé** - Les bilans thermiques et l'investigation complète de l'écoulement d'air sur une plaque d'acier noir épaisse verticale chauffée à flux de chaleur uniforme (1000 à 8000W.m$^{-2}$) sont obtenus par méthode thermoanémométrique et pyrométrie IR. Divers modèles physiques résolus au moyen d'un code de calcul commercial (FLUENT) sont ensuite comparés aux valeurs expérimentales.

**Nomenclature**

| | | | |
|---|---|---|---|
| $b$ | épaisseur de plaque chauffée, m | $U_m$ | vitesse verticale maximum, m/s |
| $d$ | distance entre microthermocouples, m | $x$ | distance verticale au bord d'attaque, m |
| $G_k$ | production turbulente motrice, kg.m$^{-1}$.s$^{-3}$ | $y$ | distance horizontale à la plaque, m |
| $Gr_x$ | nombre de Grashof local $g\beta_0(T_w\text{-}T_0)x^3/\nu_f^2$ | *Symboles grecs* | |
| $h_c$ | coefficient d'échange convectif, W.m$^{-2}$.K$^{-1}$ | $\beta$ | coefficient de dilatation, K$^{-1}$ |
| $H$ | hauteur de plaque, m | $\varepsilon$ | dissipation de $k$, m$^2$.s$^{-3}$, ou émissivité |
| $I_T$ | intensité thermique turbulente $\sqrt{\overline{T'^2}}/(T_w - T_0)$ | $\lambda$ | conductivité thermique de l'air, W.m$^{-1}$.K$^{-1}$ |
| $I_U$ | intensité de turbulence dynamique $\sqrt{\overline{u'^2}}/U_m$ | $\theta$ | température réduite $(T\text{-}T_0)/(T_w\text{-}T_0)$ |
| $k$ | énergie cinétique turbulente, m$^2$.s$^{-2}$ | $\nu$ | viscosité cinématique, m$^2$.s$^{-1}$ |
| $Nu$ | nombre de Nusselt $h_c x/\lambda_f$ | $\zeta$ | distance horizontale réduite $-y\left(\dfrac{\partial\theta}{\partial y}\right)_w$ |
| $Q,q$ | densité du flux de chaleur, W.m$^{-2}$ | *Indices* | |
| $T$ | température, °C ou K | $c$ | convectif |
| $T'$ | fluctuation de température, °C ou K | $f$ | à la température de film $(T_w+T_0)/2$ |
| $U$ | vitesse verticale, m/s | $r$ | radiatif |
| $u'$ | fluctuation de vitesse verticale, m/s | $w$ | indice relatif à la paroi |
| $U_b$ | vitesse de référence $[g\beta(T_w\text{-}T_0)x]^{1/2}$ | $0$ | milieu ambiant, ou imposé |

## 1. Introduction

Le couplage du rayonnement et de la conduction avec la convection naturelle turbulente joue un rôle important dans le bilan thermique des systèmes de chauffage tels que les fours, les échangeurs de chaleur, les panneaux solaires ou les récupérateurs de chaleur. L'efficacité de refroidissement et l'optimisation thermique de ceux-ci dépendent étroitement de la transition turbulente et des effets du rayonnement sur l'échange convectif local [1].

La paroi verticale représente à cet égard un cas fondamental qui permet une investigation approfondie de ces mécanismes et l'établissement de bilans thermiques complets. Peu d'études ont été menées dans le domaine des températures élevées sur parois verticales [2][3][4]. Au regard des enjeux industriels, l'influence des propriétés physiques réelles des parois chauffées doit être prise en compte. Nous nous attachons ici à mettre en évidence la déstabilisation de la couche limite sur de très courtes hauteurs lorsqu'elle est soumise à des flux élevés et son impact sur les transferts de chaleur. Nous avons choisi un code de calcul commercial (FLUENT) pour la simulation numérique en vue d'applications à des cas plus complexes.

## 2. Etude expérimentale

### 2.1. Dispositif expérimental et métrologie

Le système étudié (*figure 1*) est constitué d'une plaque plane en acier noir "patinable" (oxydation protectrice) utilisé dans les fours, foyers de combustion et tubages de cheminée industriels, d'épaisseur $b$ = 4 mm, de hauteur $H$ = 50 cm égale à la largeur, dont une face est chauffée à flux uniforme $q_0$ par une nappe de rubans électriques plats épais de 100μm. L'autre face de la plaque et ses bords haut et bas sont libres. Les déperditions thermiques arrière et latérales sont limitées à moins de 1% par 15 cm d'isolant thermique minéral.

Afin d'assurer un caractère essentiellement bidimensionnel à l'écoulement d'air et aux transferts thermiques, ce système est flanqué de deux murs verticaux isolants épais hauts de 3 m et larges de 1,5 m. La plaque étant fixée à 1,5 m du sol et 3 m du plafond, l'influence de ceux-ci peut être considérée comme négligeable. Les parois de la salle sont considérées comme des corps noirs. La stratification thermique verticale est très faible (0,54 K.m$^{-1}$ au maximum pour $q_0$ = 8000 W.m$^{-2}$). L'installation est détaillée dans [5].

Les températures des parois de toute la maquette et de l'air ambiant sont mesurées par 64 thermocouples multiplexés. L'émissivité totale normale de la plaque d'acier est mesurée par pyromètre 8-14μm (précision supérieure à 1%), une série de spectrogrammes par FTIR de 2,5 à 10μm ayant établi le caractère radiatif gris de sa surface, par ailleurs quasi-diffusante.

La mesure des flux convectifs pariétaux, des profils thermoconvectifs moyens et turbulents de l'écoulement d'air et des caractéristiques turbulentes est assurée par une *sonde unique à deux microthermocouples parallèles* de type K de 15 mm de long et 25 μm de diamètre, soudés au bout d'une longue canne blindée en quartz. Le positionnement des capteurs dans tout l'écoulement jusqu'à la paroi se fait à 12,5 μm près par microguidages. Un soin particulier a été apporté à l'élimination des bruits (construction d'un amplificateur d'instrumentation blindé, blindage des câbles, mise à la terre centrale, filtrage électrique des appareils). La précision des mesures de température est d'environ 0,5°C.

La distance $d$ entre microthermocouples, la fréquence d'échantillonnage, la longueur des enregistrements ont été fixés respectivement à 2,3 mm, 400 Hz et 8192 de manière à offrir une précision suffisante dans le calcul de la vitesse par mesure des temps de transit des fluctuations thermiques entre les deux capteurs, tout en limitant la durée totale de l'investigation de l'écoulement à 10 heures (1296 enregistrements). Nous avons opté pour une valeur aussi petite que possible de $d$ pour conserver l'intégralité de l'information contenue dans la signature thermique des turbulences, leur spectre s'élargissant au-delà de 150Hz à 8000 W.m$^{-2}$ d'après nos essais. Diverses études ont montré qu'en dehors de la sous-couche visqueuse où elle augmente fortement, la vitesse "convective" de transport de ces fluctuations $U_c$ est pratiquement confondue avec la vitesse de l'air $U$ [6], mais que la décorrélation des fluctuations thermiques s'accroît rapidement avec leur fréquence [7].

La mesure de la vitesse $U_c$ et de son intensité de turbulence $I_U$ est obtenue par l'algorithme de *thermoanémométrie par intercorrélation à fenêtre glissante* (TIFG) que nous avons développé, dont l'étude paramétrique est présentée dans [8].

La validation de la thermoanémométrie par intercorrélation (CCV)[9] dans un écoulement libre de convection naturelle conclut à une précision meilleure que 5%, ce que confirment nos mesures comparées à [4]. Ces dernières montrent en revanche que $U_c$ augmente fortement dans la sous-couche visqueuse ($y < 2$mm). Un filtrage passe-bas de Fourier dans la sous-couche visqueuse conduit à des valeurs satisfaisantes, avec un profil en $y^3$ à la paroi (*figure 6*).

## 2.2. Résultats et interprétation

La densité $q_c(x)$ du flux convectif et le coefficient d'échange $h_c(x)$ sont calculés le long de la plaque par régression linéaire des profils thermiques dans la sous-couche visqueuse, la densité du flux radiatif $q_r(x)$ étant directement obtenue à partir des mesures des émissivités $\varepsilon_r$ et des températures pariétales $T_w$. Les profils $T_w(x)$ et $h_c(x)$ sont donnés en *figures 2 et 3*. L'intégration de $q_c$ et $q_r$ sur la face libre et les bords de la plaque conduit aux bilans thermiques du *tableau 1* du §3.3 (précision< 3% sur $h_c$, 1% sur $q_r$; bilans bouclés à 1% près).

On observe que *les échanges radiatifs dominent largement* pour $q_0 > 2000$ W.m$^{-2}$ et atteignent 76,7% à 8000 W.m$^{-2}$ avec le profil d'émissivité mesuré ($\varepsilon_r$ varie entre 0,698 et 0,965). A 2000 W.m$^{-2}$, des bouffées de fluctuations thermiques sinusoïdales sont enregistrées à partir de 32,5cm et la décroissance laminaire de $h_c$ est stoppée. Cette *déstabilisation de la couche limite* est plus marquée à 4000 W.m$^{-2}$ où les bouffées sinusoïdales s'amplifient au-dessus de 30 cm en prenant un caractère chaotique dès 38 cm, $h_c$ restant presque constant à 4 W.m$^{-2}$.K$^{-1}$ avant d'augmenter en haut de plaque [5]. A 8000 W.m$^{-2}$, la *transition turbulente* débute vers 12,5 cm et les fluctuations thermiques sont quasi-turbulentes à partir de 25 cm. Le profil $h_c(x)$ est alors irrégulier (forte interaction avec le rayonnement) et remonte rapidement en haut de plaque. Si la corrélation du flux de chaleur $Nu_x(Gr_x)$ suit bien la corrélation laminaire [10] jusqu'à $Gr_x = 10^8$, le *régime de transition* est bien représenté par

$$Nu_x = 0{,}179\, Gr_x^{0{,}273} \qquad (1)$$

avec une diminution de $Nu_x$ (effet de la variation des propriétés physiques de l'air à température élevée) et une pente plus élevée qu'en régime laminaire (*figure 4*).

A 8000 W.m$^{-2}$, la corrélation remonte largement au-dessus en début de zone de turbulence développée dès $Gr_x = 10^8$. Au-delà de $Gr_x = 3.10^8$, elle présente les perturbations prévisibles que nous attribuons à la forte prédominance du rayonnement qui perturbe, en l'affaiblissant, la densité du flux de chaleur traversant la sous-couche motrice, accroissant la déstabilisation de la couche limite et la brusque remontée de $Nu_x$ au-delà de $Gr_x = 10^9$.

Le *nombre de Grashof critique* $Gr_c$, repérant le début d'amplification des fluctuations, décroît fortement sous $10^9$ ($10^8$ puis $2.10^7$ lorsque $q_0$ passe de 2000 à 8000 W.m$^{-2}$ [2]).

Les *figures 5 et 6* donnent les profils réduits $\theta(\zeta)$ et $U_c/U_m(\zeta)$ dans la zone de turbulence développée à 8000 W.m$^{-2}$ (*cf.* [5][8]). Ils suivent assez étroitement les résultats de [4] obtenus par LDA et microthermocouple à $Gr_x = 3{,}4.10^{11}$ sur une paroi à environ 430°C. En particulier, la vitesse maximale est $U_m = 0{,}707$m/s en $\zeta = 1{,}3$ avec $U_m/U_b = 0{,}328$ (0,31 en $\zeta = 1{,}4$ dans [4]). Les graphes $I_T(\zeta)$ et $I_U(\zeta)$ non reproduits ici [8] sont également proches de ceux de [4], avec un maximum de $I_T$ de 0,127 à $\zeta = 0{,}58$ (resp. 0,14 et 0,5 dans [4]). Le graphe $I_U(\zeta)$, plus dispersé du fait de la faible longueur des enregistrements, est à peu près uniforme autour de 0,3 dans la sous-couche externe et décroît vers 0,15 dans la sous-couche "motrice" (tampon).

## 3. Modélisation et simulation numérique

### 3.1. Conditions aux limites

Le modèle prend en compte les dimensions et propriétés physiques réelles du système expérimental, et les propriétés physiques thermovariables de l'air. Les densités du flux de chaleur appliqué sont $q_0$ = 2000, 4000 et 8000 W.m$^{-2}$; l'émissivité est uniforme ($\varepsilon_r$ = 0,5; 0,66; 0,88; 1) ou suit le profil réel. Les limites imposées aux frontières correspondent aux dimensions réelles de la salle (hauteur 5 m; longueur 10 m), avec deux types de conditions:
- enceinte fermée à parois isothermes considérées comme corps noirs à 293K
- frontières ouvertes isothermes à 293K, à vitesses normales aux quatre frontières.

### 3.2. Modèles physiques et résolution numérique

La fermeture des équations de transport moyennées de Reynolds est effectuée selon des modèles turbulents en $k$-$\varepsilon$ avec production d'énergie cinétique turbulente $G_k$ par poussées d'Archimède formulée selon une hypothèse de diffusion simple de gradient. Après application du modèle $k$-$\varepsilon$ standard avec lois de parois standard [5], les résultats obtenus sont comparés à ceux issus d'un modèle $k$-$\varepsilon$ basé sur la théorie du *Groupe de Renormalisation* (RNG) [11] pour prendre en compte les effets des parois sur les échelles de la turbulence à bas nombre de Reynolds turbulent. L'abandon des lois de paroi standard au profit d'un *modèle zonal à deux couches* [12] est aussi testé; il permet de fixer des conditions aux limites pour $\varepsilon$ plus réalistes dans les zones à bas Reynolds, de sorte que $k$ tende progressivement vers 0 aux parois.

Les échanges radiatifs sont calculés par *méthode des transferts discrets radiatifs* (DTRM) [13], rapide et précise en milieu peu absorbant et peu diffusif comme l'air, et bien adaptée à la *méthode des volumes finis* [14] retenue pour la résolution des équations de l'écoulement. L'angle de discrétisation choisi pour le rayonnement est de 5° (précision des bilans thermiques de 1%).

La grille fine choisie pour discrétiser le domaine de calcul comporte 178 cellules horizontalement (50 dans la couche limite avec resserrement à la paroi, dont 2 à 5 dans la sous-couche visqueuse) et 142 verticalement (dont 50 le long de la plaque, uniformément réparties). Une *résolution multigrille* est utilisée pour accélérer la convergence. La discussion de ces choix, des modèles physiques, des méthodes de relaxation est exposée dans [5].

### 3.3. Résultats comparés

Le *tableau 1* récapitule les grandeurs principales à $q_0$ = 8000 W.m$^{-2}$. Avec les profils thermiques $T_w/T_0$, il montre que *le modèle le plus proche des valeurs expérimentales est le modèle RNG-k-$\varepsilon$ avec modèle zonal à deux couches appliqué à un domaine ouvert et une paroi à profil d'émissivité réel*. Les profils $h_c(x)$ et $Nu_x$ restent cependant supérieurs à ceux observés, même si la pente 0,269 de $Nu_x(Gr_x)$ est proche de la valeur expérimentale 0,273. Le profil $\theta(\zeta)$ est un peu faible dans la sous-couche externe. Le profil $U_c/U_m(\zeta)$ dépasse le profil réel et présente une déflexion d'écoulement à partir du bord d'attaque non observée expérimentalement. Nous attribuons ces écarts à une dissipation $\varepsilon$ trop faible (terme $G_k$ et $\varepsilon_w$ de forme discutable dans FLUENT [15]), exagérant la production de $k$ au bord d'attaque; à la négligence des fluctuations de $\rho$ et à une surestimation des pertes dans l'isolant.

L'ouverture du domaine de calcul a un effet prépondérant et permet de supprimer les contraintes de la bidimensionnalité en cavité fermée qui génère des recirculations artificielles accroissant fortement la turbulence, la vitesse et l'échange convectif sur la plaque. Le modèle zonal à deux couches offre un bon compromis en forçant $k$ à s'annuler vers les parois.

Tableau 1: *Comparaison des modèles numériques et de l'expérience ($q_0 = 8000$ W.m$^{-2}$)*

| domaine | modèle | parois | $\varepsilon_r$ | $Q_r/Q_T$ (%) | $T_{w\,max}$ (K) | $U_m$ (m.s$^{-1}$) | $h_{c\,min}$ (W.m$^{-2}$.K$^{-1}$) |
|---|---|---|---|---|---|---|---|
| Fermés | $k$-$\varepsilon$ | standard | 0,88 | 45,3 | 542,9 | 1,19 | 14,01 |
| | RNG $k$-$\varepsilon$ | zonal | 0,88 | 63,5 | 582,7 | 1,20 | 6,91 |
| | RNG $k$-$\varepsilon$ | zonal | var. | 64,6 | 585,7 | 1,18 | 7,60 |
| Ouverts | $k$-$\varepsilon$ | standard | 0,88 | 62,4 | 581,3 | 1,00 | 8,90 |
| M3 | RNG $k$-$\varepsilon$ | standard | 0,88 | 65,2 | 586,9 | 0,990 | 7,83 |
| M2 | RNG $k$-$\varepsilon$ | zonal | 0,88 | 71,1 | 596,5 | 1,01 | 5,92 |
| M1 | RNG $k$-$\varepsilon$ | zonal | var. | 71,2 | 600,4 | 1,01 | 5,82 |
| | expérience | | var. | 76,7 | 611,7 | 0,707 | 4,73 |

## 4. Conclusions

L'étude des transferts thermiques et des écoulements d'air en convection naturelle le long d'une plaque conductrice rayonnante fortement chauffée montre une déstabilisation rapide de la couche limite, attribuée au rayonnement et à la redistribution des flux conductifs dans la plaque, abaissant le nombre de Grashof critique et augmentant l'échange convectif en haut de plaque. La thermoanémométrie par intercorrélation à fenêtre glissante (TIFG) constitue une méthode précise et efficace de détermination des caractéristiques moyennes et turbulentes des écoulements de convection naturelle et des transferts thermiques à haute température.

L'utilisation d'un modèle bidimensionnel RNG-$k$-$\varepsilon$ associé à un modèle zonal bi-couche de la turbulence et à un modèle radiatif de transfert discret reproduit convenablement les résultats expérimentaux si les frontières du domaine de calcul sont ouvertes, mais surestime légèrement l'échange convectif, les vitesses et l'effet des bords d'attaque.

**Références**

[1] R. Siegel, J.R. Howel, Thermal radiation heat transfer, HPC, Washington, 1992
[2] D.L. Siebers, R.F. Moffat, R.G. Schwind, Experimental, variable properties natural convection from a large, vertical, flat surface, J. Heat Transfer 107 (1985) 124-132
[3] L.R. Cairnie, A.J. Harrison, Natural convection adjacent to a vertical isothermal hot plate with a high surface-to-ambiant temperature difference, Int. J. Heat Mass Transfer 25 (7) (1982) 925-934
[4] S. Kato, S. Murakami, R. Yoshie, Experimental and numerical study on natural convection with strong density variation along a heated vertical plate, 9$^{th}$ Symp. Turb. Shear Flows, Kyoto, Japon, 1993, pp. 12.5.1-12.5.6
[5] T. Delarochelambert, Etude expérimentale et théorique des transferts thermiques couplés en convection naturelle à travers une double paroi verticale à haute densité de flux de chaleur, Thèse, Université de Haute-Alsace, 1997
[6] T. Tsuji, Y. Nagano, M. Tagawa, Experiment on spatio-temporal turbulent structures of a natural convection boundary layer, J. Heat Transfer 114 (1992) pp. 901-908
[7] R. Cheesewright, A. Dastbaz, The structure of turbulence in a natural convection boundary layer, Proc. 4$^{th}$ Turb. Shear Flows Symp., Karlsruhe, Allemagne, 1983, pp. 17.25-17.30
[8] T. Delarochelambert, G. Prado, Thermoanémométrie corrélative en convection naturelle, IV$^{e}$ Colloque Interuniversitaire Franco-Québécois, Montréal, 1999, pp. 17-22
[9] V. Motevalli, C.H. Marks, B.J. McCaffrey, Cross-correlation velocimetry for measurement of velocity and temperature profiles in low-speed, turbulent, nonisothermal flows, J. Heat Transfer 114 (1992) 331-337

[10] A. Pirovano, S. Viannay, M. Jannot, Convection naturelle en régime turbulent le long d'une plaque plane verticale, 9th Int. Heat Transfer Conf., 1970, (4) NC 1.8, pp. 1-12
[11] V. Yakhot, S.A. Orszag, Renormalization group analysis of turbulence. I. Basic theory, J. Sci. Comp. 1 (1986) 3-51
[12] H.C. Chen, V.C. Patel, Near-wall turbulence models for complex flows including separation, Am. Int. Aer. Astr. J. 26 (6) (1988) 641-648
[13] F.C. Lockwood, N.G. Shah, A new radiation solution method for incorporation in general combustion prediction procedures, 18th Int. Symp. Comb., Waterloo, Canada, 1980, pp. 1405-1414
[14] S.V. Patankar, Numerical heat transfer and fluid flow, HPC, New York, 1980
[15] W.P. Jones, B.E. Launder, The prediction of laminarization with a two-equation model of turbulence, Int. J. Heat Mass Transfer 15 (1972) 301-314

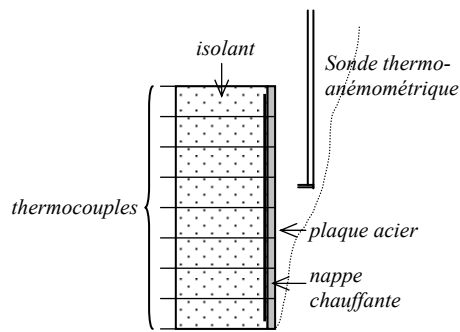

Figure 1 : *Dispositif expérimental*

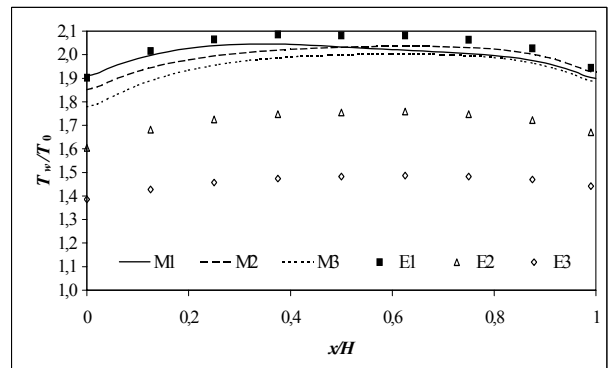

Figure 2 : *Température pariétale [modèles M1,2,3 à $8000 W.m^{-2}$; Exp.: 1,2,3 (8000, 4000, 2000 $W.m^{-2}$)]*

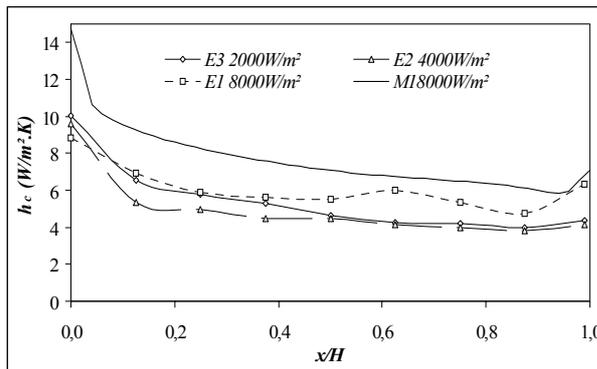

Figure 3 : *Coefficient d'échange convectif [modèle M1 à $8000 W.m^{-2}$; Exp.E1,E2,E3]*

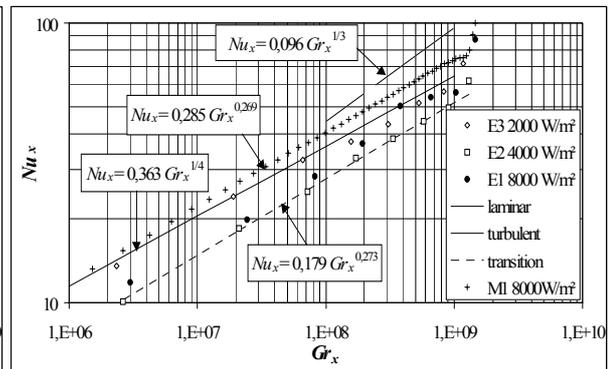

Figure 4 : *Corrélations du flux de chaleur [modèle M1 à $8000 W.m^{-2}$; Exp.E1,E2,E3]*

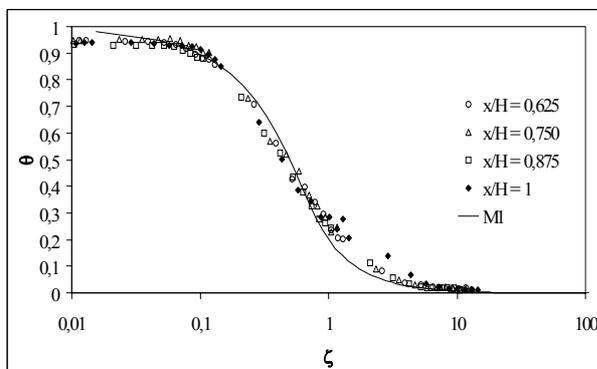

Figure 5 : *Profil thermique convectif réduit à $8000 W.m^{-2}$ [M1 : x/H=1]*

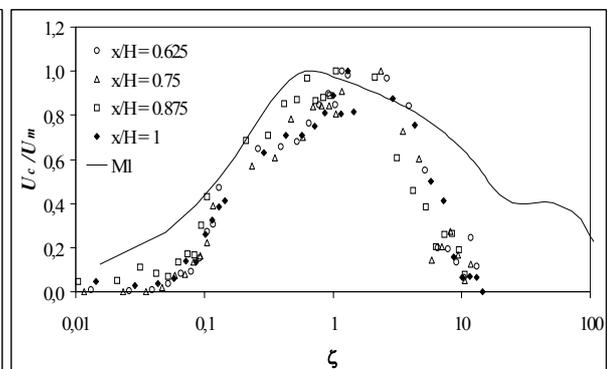

Figure 6 : *Profil dynamique convectif réduit à $8000 W.m^{-2}$ [M1 : x/H=1]*